\documentclass[prl,twocolumn,aps,showpacs,10pt]{revtex4}
\usepackage{graphicx}
\usepackage{bm}
\usepackage{amssymb} 
\usepackage{color}

\begin{document}

\title{Localized states  due to expulsion of resonant impurity levels from the continuum in bilayer graphene}

\author{V. V. Mkhitaryan and E. G. Mishchenko}

\affiliation{Department of Physics and Astronomy, University of Utah, Salt Lake
City, UT 84112, USA}

\begin{abstract}
Anderson impurity problem is considered for a graphene bilayer
subject to a gap-opening bias. In-gap localized states are produced
even when the impurity level overlaps with the continuum of band
electrons. The effect depends strongly on the polarity of the
applied bias as long as hybridization with the impurity occurs
within a single layer. For an impurity level inside the conduction
band a positive bias creates the new localized in-gap state. A
negative bias does not produce the same result and leads to a simple
broadening of the impurity level. The implications for transport are
discussed including a possibility of gate-controlled Kondo effect.
\end{abstract}

\maketitle

{\it Introduction}. Electronic properties of graphene are affected
by short-range scatterers (for the review see Ref.~\cite{cn}).
Remarkably, these properties significantly depend on the character
of disorder \cite{Mirlin06}. It has been demonstrated that the limit
of {\it strong} scattering is realized via resonances occurring for
vacancies \cite{PGC06, PGLPC06, PLC08} and adsorbed atoms
\cite{WYLGK10, Robinson,  SPG07, WKL09, Mucciolo}. Such resonances
lead to the modification of the bare spectrum near the charge
neutrality point and strongly affect the transport properties.

Moreover, the localized impurity levels in a graphene host could
potentially result in a formation of local magnetic moments
\cite{Kotov08, UYTPC09, URC11, Ting}.  This situation is of a
special interest as it opens up a perspective of controlled
switching of the local moments by gating. As far as the transport
properties are concerned the local magnetic moments famously lead to
the Kondo effect. The band structure of the monolayer graphene makes
this effect different than that in normal metals \cite{Baskaran08,
Fradkin96, Vojta10}. On one hand, narrowing of the impurity level
due to the vanishing density of states (DOS) in graphene, $\rho(E_F)
\propto |E_F|^\alpha$, facilitates the formation of a local moment
near the charge neutrality point. On the other hand, the low DOS
disfavors the Kondo resonance and results in the  suppression of the
Kondo temperature when $E_F\to 0$. In case of non-interacting
electrons, $\alpha=1$, the suppression is power-law. If the effects
of electron-electron renormalization of the spectrum are included,
so that $\alpha<1$, the Kondo temperature is suppressed
exponentially.

The theory of resonant impurity states in {\it bilayer} graphene has
received much less attention despite its greater technological
potential. The most important property of bilayer graphene to this
extent is its gap tunable by external bias \cite{KCM,ZTG,MLS}. It is
therefore important to elucidate how strongly the gap is affected by
the disorder, which generally tends to smear the gap. For
short-range impurity potential  there exist ``tail-states'' with DOS
decaying exponentially from the band edge \cite{NCN07,DB08,MR08}.
Numerical studies also indicate the onset of well-defined midgap
peaks in the presence of resonant impurities \cite{Mucciolo}.

\begin{figure}[h]
\centerline{\includegraphics[width=80mm,angle=0]{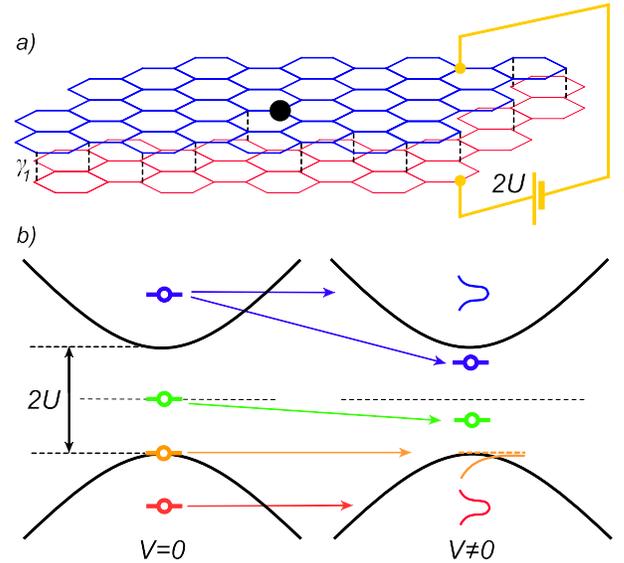}}
\caption{a) Bilayer graphene subject to the gap-opening bias $2U$.
The resonant impurity (black circle) is hybridized with
$\pi$-electrons of a carbon atom in the upper layer. Tunneling
$\gamma_1$ is indicated by the vertical dashed lines. b) Band
structure of the system for different impurity energies
$\varepsilon_0$. The left picture corresponds to a zero
hybridization amplitude $V$. The right one illustrates the
modification of the spectrum for finite couplings $V$ and positive
bias $U>0$. The level with $\varepsilon_0>U$ (blue) broadens due
to coupling to the band electrons while generating an additional
in-gap state. The level originally positioned inside the gap
(shown with green for a specific case $\varepsilon_0=0$) remains
localized but is renormalized towards lower energies. States
overlapping with the valence band $\varepsilon_0<-U$ (red) are
broadened without generating any additional in-gap states. The
level at the top of the valence band (orange) leads to a
square-root singularity in the density of states. } \label{fig1}
\end{figure}

In the present paper we analyze the effect of the Anderson impurity
hybridized with the band electrons of bilayer graphene host. We
demonstrate that a single impurity level with energy $\varepsilon_0$
induces a localized in-gap state. This is a rather trivial statement
whenever $\varepsilon_0$ falls within the (gate-induced) bandgap.
More interestingly, we find that the in-gap state appears even if
the impurity level sits inside the {\it continuum} of itinerant
states, i.e. above or below the bandgap, depending on the {\it
polarity} of the applied interlayer bias. We consider a realistic
bilayer setup, shown in Fig.~1, of an impurity sitting on top of the
upper layer. Consider the case of a positive applied bias $2U/e$ as
shown in the picture ($-e$ is the electron charge). We emphasize
that in this case the localized state appears {\it no matter how
high} the impurity level $\varepsilon_0$ is above the charge
neutrality point. To the contrary, the impurity state overlapping
with the valence band, $\varepsilon_0<-U$, does not generate a
localized state. It simply acquires a finite width (and the energy
renormalization) from the interaction with the valence band states.

The localized states of the similar nature  that appear due to
hybridization with an impurity level are known to occur in various
systems. In a one-dimensional band with a finite bandwidth a pair of
states split off near both band edges \cite{Mahan}. In conventional
$s$-wave superconductors similar bound states are long known
\cite{MachShib72, Shiba73}. However, they lie extremely close to the
superconducting gap (typically within $\sim 10^{-3}$ of its width),
and thus are hardly discernable against the divergent
superconducting DOS. In contrast, in bilayer graphene it is possible
to ``place'' the localized state virtually {\it anywhere} within the
bandgap simply by controlling the applied bias. In particular, this
can lead to the emergence of a local magnetic moment and a formation
of the Kondo resonance, more robust than in a monolayer, even when
the bare impurity level $\varepsilon_0$ is {\it above} the Fermi
level.

{\it Anderson impurity in bilayer graphene.} The low-energy
Hamiltonian of bilayer graphene is a $4\times4$ matrix in
layer/sublattice space \cite{MCF}. If in addition the energies of
interest are less than the interlayer tunneling rate the Hamiltonian
could be reduced further to the $2\times2$ matrix acting on the
layer index. In the presence of a gap-opening electrostatic
potential difference $2U$ such Hamiltonian has the form
\cite{MCF,cn},
\begin{eqnarray}\label{LEHam}
\hat H({\bf p})
=\frac{1}{2m}\left([p_x^2-p_y^2]\hat\sigma_x+2p_xp_y\hat\sigma_y
\right)+U\hat\sigma_z,
\end{eqnarray}
where $\hat \sigma_i$ are the Pauli matrices, and ${\bf p}$ is the
electron quasimomentum. The effective mass $m=\gamma_1/2v^2=0.03m_0$
is determined by the interlayer coupling, $\gamma_1=0.35$ eV, and
the Dirac velocity, $v=10^6$ m/s. This Hamiltonian acts on the
two-component wavefunction, ${\hat \psi}^\dagger_{\bf
p}=(\psi^\ast_{1 \bf p}, \psi^\ast_{2 \bf p})$, with $ \psi_{\sigma
\bf p}$ referring to the amplitudes on upper ($\sigma=1$) and lower
($\sigma=2$) layers.  The total Hamiltonian of the system in the
second quantization representation is
\begin{eqnarray}\label{SysHam}
{\cal H}=\! \sum_{\bf p}{\hat \psi}^\dagger_{\bf p}\hat H({\bf p})
{\hat \psi}_{\bf p}+\varepsilon_0 d^\dagger d+ V\! \sum_{\bf p}
(\psi^\dagger_{1 \bf p}d+d^\dagger \psi_{1\bf p}),
\end{eqnarray}
where the second term describes the localized level and the last
term defines the hybridization of this level with band-electrons
in the upper layer only. This choice, while not of principal
importance for the calculations, ensures the asymmetry with
respect to the polarity of $U$ as we show below. The properties of
the localized electrons are most conveniently described by its
Green's functions,
\begin{equation}
D(t)=-i\langle T d(t) d^\dagger (0) \rangle.
\end{equation}
This function satisfies the equation which follows from
Eq.~(\ref{SysHam}). In the energy representation,
\begin{eqnarray}
\label{local_eq} (E-\varepsilon_0) D(E)= 1+V {\cal G}_{1}(0,E).
\end{eqnarray}
The mixed Green's function in the latter equation is defined
according to
\begin{equation}
{\cal G}_{\sigma}({\bf r},t)=-i\langle T \psi_{\sigma}({\bf r},t)
d^\dagger(0) \rangle
\end{equation}
and satisfies the equation
\begin{eqnarray}
\label{mixed_eq} (E-\hat H)_{\sigma\sigma'} {\cal G}_{\sigma'} ({\bf
r},E)=
V\delta({\bf r})\delta_{\sigma 1}D(E). 
\end{eqnarray}
The two equations (\ref{local_eq}) and (\ref{mixed_eq}) form a
closed system that readily yields the solution,
\begin{equation}\label{DE}
D(E)=\frac{1}{E-\varepsilon_0 - V^2G^{(0)}_{11}(E)},
\end{equation}
where we introduced the Green's function of free band electrons, $
\hat G^{(0)}(E)= \sum_{{\bf p}} [E-\hat H({\bf
p})+i\eta~\text{sign}E]^{-1}$. In what follows we are mostly
interested in the gapped region, $-U<E<U$, where  after simple
calculations \cite{valleys} we obtain
$G^{(0)}_{11}(E)=-\frac{m}{2}\frac{E+U}{\sqrt{U^2-E^2}}$.
Remarkably, in this region there exists a bound state which emerges
as a simple pole of the Green's function $D(E)$, Eq. (\ref{DE}). Its
energy satisfies the relation
\begin{equation}\label{pole}
E-\varepsilon_0+\Gamma \sqrt{\frac{U+E}{U-E}}=0.
\end{equation}
Here we introduced the parameter, $\Gamma=mV^2/2$, whose meaning is
the inverse lifetime of the original impurity level high above the
gap, $E\gg U$. Remarkably, the equation (\ref{pole}) is asymmetric
with respect to the sign of the applied bias $U$. Let us concentrate
on the case of positive bias $U>0$ below. The generalization to the
opposite case is almost evident and will be discussed later. It is
easy to see that Eq. (\ref{pole}) has a true in-gap bound state
solution, $\widetilde \varepsilon$, for {\it any} impurity level
such that $\varepsilon_0> -U$. Fig.~\ref{fig1} illustrates the
energy spectrum of the system for various values of $\varepsilon_0$.
The most trivial case is realized when $\varepsilon_0$ falls within
the bandgap $-U<\varepsilon_0<U$. Here the energy of the resulting
localized state is lowered by the hybridization, $\widetilde
\varepsilon<\varepsilon_0$. Moreover, impurity level inside the
positive energy band $\varepsilon_0>U$ {\it always} creates an
in-gap localized state, regardless of how large $\varepsilon_0$
and/or how narrow the gap is.

\begin{figure}[t]
\centerline{\includegraphics[width=90mm,angle=0]{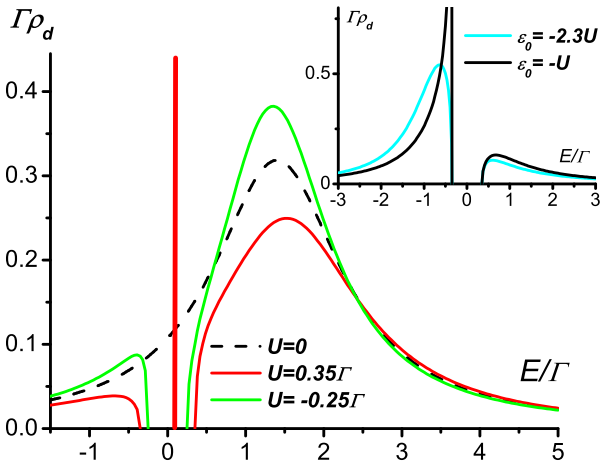}}
\caption{a) Dimensionless density of states (scaled with the
coupling strength $\Gamma$) of the $d$-electrons plotted in units
of $1/\Gamma$ for $\Gamma=\gamma_1/2$,
$\varepsilon_0=0.7\gamma_1$, and three different values of bias
$U$: zero bias (black dashed line), positive bias (red), negative
bias (green). The bound in-gap state appears as a sharp peak for
$U>0$, but not for $U<0$. Inset: development of a square-root
singularity when the impurity level is approaching the top of the
valence band from the inside of it.} \label{densdis}
\end{figure}

The effect of the impurity level is two-fold. First, its interaction
with band electrons broadens its density of states. This could be
readily seen from Eq.~(\ref{DE}) with $|E|>U$, where the self energy
of the $d$-state becomes purely imaginary,
$G^{(0)}_{11}(E)=-i\frac{m}{2} \frac{E+U}{\sqrt{E^2-U^2}}$. The
density of states of the impurity level $\rho_d(E)$ is modified from
the simple Lorentzian, which is the case when $U=0$:
\begin{equation}
\rho_d(E)=-\frac{\Im D_R(E)}{\pi}=\frac{\Gamma\sqrt{E^2-U^2}/\pi}
{(E-\varepsilon_0)^2|E-U|+\Gamma^2|E+U|}.
\end{equation}
Fig.~\ref{densdis} shows the density of impurity states for
different values of the gap. The second feature is the appearance of
the delta-function peak, corresponding to the new localized state,
Eq.~(\ref{pole}), also indicated in Fig.~\ref{densdis}. The total
area under any curve is $\int\rho_d(E)dE=1$. In evaluating this area
one notices that the in-gap peak in $\rho_d(E)$ has a spectral
weight smaller than unity:
\begin{equation}\label{weighted}
\rho_d(E)=
\frac{(U-\widetilde\varepsilon)^{3/2}(U+\widetilde\varepsilon)^{1/2}}
{(U-\widetilde\varepsilon)^{3/2}(U+\widetilde\varepsilon)^{1/2}+
U\Gamma}\, \delta(E-\widetilde\varepsilon).
\end{equation}
The remaining spectral weight that completes the total weight of the
in-gap state to the unity comes from the band electrons. This can be
verified by calculating the band electron Green's function,
$G_{\sigma\sigma'}({\bf r},{\bf r'},t)=-i\langle T \psi_\sigma({\bf
r},t) \psi^\dagger _{\sigma'}({\bf r'},0) \rangle$, accomplished by
constructing a system of coupled equations similar to
Eqs.~(\ref{local_eq}) and (\ref{mixed_eq}).

{\it Impurity levels overlapping with the valence band}. When the
impurity level resides inside the lower continuum of band states,
$\varepsilon_0<-U$, the broadening of the level {\it is not}
accompanied by the creation of an in-gap state, see Fig.~\ref{fig1}.
Far from the band edge $-\varepsilon_0 \gg U$ the lineshape is a
simple Lorentzian but is distorted as $\varepsilon_0$ moves closer
to the edge. At $\varepsilon_0=-U$ the density of states develops a
singularity $\rho_d \sim \Gamma^{-1}\sqrt{U/|E+U|}$. Both these
cases are shown in the inset on Fig.~\ref{densdis}.

The asymmetry between the positive and negative energies
$\varepsilon_0$ and the sensitivity to the sign of the external bias
is a manifestation of breaking of the symmetry between the two
layers by the impurity position. If the impurity particle is
hybridized equally with both monolayers the symmetry is restored and
the in-gap bound state is always present. It is found from an
equation similar to Eq.~(\ref{pole}), where the square root is
replaced with $2E/\sqrt{U^2-E^2}$.

{\it $4 \times 4$ bilayer Hamiltonian.} In the previous discussion
we utilized the simplified two-band model, which is usually
justified when the typical energies are small compared with the
interlayer tunneling, $E,~U \ll \gamma_1$. We are now going to
discuss the new qualitative features that the inclusion of all four
$\pi$-electron bands leads to.  The full four-band Hamiltonian is a
$4\times 4$ matrix \cite{MCF},
\begin{equation}
\label{fullHam} \hat H({\bf p})=\left(\begin{array}{cccc} U & vp_+ &
0 & 0
\\vp_- & U & \gamma_1 & 0\\ 0 & \gamma_1 & -U & vp_- \\ 0 & 0& vp_+
& -U
\end{array} \right),
\end{equation}
where $p_{\pm}=p_x\pm ip_y$. The first two rows/columns of Eq.
(\ref{fullHam}) refer to the two sublattices in the upper graphene
layer and the last two to the lower layer. The formal expression for
the total Hamiltonian of the system, Eq.~(\ref{LEHam}), remains
unchanged with the exception of the dimensionality of spinors $\hat
\psi_{\bf p}$. The effect of hybridization is the strongest when the
impurity sits on a carbon atom belonging to the {\it first}
sublattice -- the one not directly coupled by tunneling $\gamma_1$
to the second layer. A rather peculiar situation of an impurity
residing on the other sublattice of the first layer is discussed
below.

The pole of Eq.~(\ref{DE}) still defines the bound state with the
corresponding self-energy given by the integral of the Green's
function over momentum,
\begin{widetext}
\begin{equation}
\label{g11} G_{11}^{(0)}(E)= \frac{U-E}{8\pi v^2}
\ln{\frac{D^4}{(U^2-E^2)(\gamma_1^2+U^2-E^2)}} - 
\frac{\gamma_1^2(U+E) +2UE(U-E)} {4\pi
v^2\gamma_1U\sqrt{1-E^2/{\widetilde U}^2}
} \left(\frac{\pi}{2}+\tan^{-1}{\frac{E^2+U^2}
{\gamma_1U\sqrt{1-E^2/{\widetilde U}^2}}} \right).
\end{equation}
\end{widetext}
One of the consequences of the expanded Halimtonian (\ref{fullHam})
is that the true gap $\widetilde U
=\gamma_1U/\sqrt{\gamma_1^2+4U^2}$ is found at a finite momentum.
Since $\widetilde U$ and is slightly less than $U$, the logarithmic
term in Eq.~(\ref{g11}) is cut-off and represents a non-significant
renormalization of $\varepsilon_0$. The most significant difference
obtained in the $4\times 4$ model is the behavior at $E\sim
-\widetilde U$, where $G_{11}^{(0)}(E)$ changes sign and therefore
{\it allows} the in-gap state solution expelled from the inside of
the {\it valence} band, $\varepsilon_0<-U$, situation forbidden by
Eq.~(\ref{pole}). Such states are residing very close to the top of
the band as the overall effect is rather weak. For the binding
energy we obtain the expression,
\begin{equation}
\label{binding11} \widetilde \varepsilon+\widetilde U =\frac{8
\Gamma^2 U^5}{\gamma_1^4 (|\varepsilon_0|-\widetilde U)^2},
\end{equation}
applicable as long as $|\varepsilon_0|-\widetilde U\gg \Gamma
U^2/\gamma_1^2$.

Let us now turn to the impurity sitting on top of an atom belonging
to the {\it second} (directly coupled by tunneling to the bottom
layer) sublattice. In this case the self-energy in Eq.~(\ref{DE}) is
given by $G_{22}^{(0)}(E)$ instead of $G_{11}^{(0)}(E)$. The former
is determined by expression similar to Eq.~(\ref{g11}) where the
term $\gamma_1^2(U+E)$ in the numerator of the second term is set to
zero. There are two notable consequences of this change. First, all
bound solutions for the impurity overlapping with the conduction
band, $\varepsilon_0>U$, completely {\it disappear}. The only
non-trivial states are those emerging from the valence band. Second,
such states are expelled stronger than in the case considered in the
previous paragraph. This is reflected in the increased binding
energy, which is given by Eq.~(\ref{binding11}) multiplied by the
extra factor $4$.

{\it Gate-controlled Kondo effect via formation of a local magnetic
moment.} A direct inclusion of the spin degree of freedom in the
Hamiltonian (\ref{SysHam}) does not change the above physical
picture unless the electron-electron interaction is taken into
account. In the presence of the spin degeneracy, this interaction is
often described by the on-site Hubbard repulsion, which under some
conditions can cause a magnetic instability \cite{Hewson}. Such a
description, however, might not be adequate for the in-gap states
found above. Indeed, as evident from Eq.~(\ref{weighted}), a
significant fraction of the spectral weight comes from the
evanescent band states, in particular, when $\widetilde \varepsilon$
is close to the band edges, $\pm U$. We consider the most intriguing
situation of an impurity level high in the conduction band
$\varepsilon_0 \gg U$, so that the in-gap state is close to $U$. To
estimate the magnitude of the Coulomb charging energy, we note that
the radius of this  state is $R\sim[2m(U-\tilde
\varepsilon)]^{-1/2}$. In turn, the binding energy can be found from
Eq. (\ref{pole}) to be $U-\tilde \varepsilon\simeq 2 \Gamma^2
U/(\varepsilon_0-U)^2$. Therefore the charging energy is $E_c\simeq
e^2/R\simeq e^2\Gamma \sqrt{mU}/\varepsilon_0$. The condition for
the formation of a well-defined magnetic moment in the $n$-doped
bilayer (with the Fermi energy $E_F>U$) is that the double occupancy
of the in-gap state is forbidden, $E_c>E_F-U$. In terms of the
gap-opening bias, this happens when $U>
E_F^2\varepsilon_0^2/E_B\Gamma^2$, where we introduce the Bohr
energy, $E_B\approx 1.47$ eV. Under these conditions the
gate-controlled Kondo effect can potentially be observed in bilayer
graphene, where one has the additional benefit of a finite DOS, in
contrast to a monolayer.

{\it Conclusion}. An important difference between the above found
resonant bound states with the previously predicted bound states due
to short-range impurities \cite{NCN07, DB08} or vacancies
\cite{CLSV10} is that in our case the emerging states can reside
anywhere inside the gap, not just close to the band edges. Wide
range for the hybridization energies for a number of impurities (H,
CH$_3$, OH, F) has been reported from the first principles from
$W\sim 0.7$ eV in Ref.~\onlinecite{WKL09} to $W\sim 5$ eV in
Ref.~\onlinecite{Robinson}.  Within the effective Dirac description
this corresponds to $V\sim (3^{3/4}a/2\hbar)W$, where the
interatomic distance $a=1.42$ \AA. Using this estimate, $\Gamma\sim
10-250$ meV. Due to this large value of $\Gamma$, the depth of the
induced in-gap bound state is comparable with $U$ for values of
$\varepsilon_0$ as large as a few $\Gamma$. In addition to the above
discussed Kondo effect, the potential experimental implications
include sub-gap gate-dependent optical absorption, as well as the
variable-range hopping transport via the impurity band, which would
form in the presence of a finite density of impurities.

Useful discussions with M.~Raikh and O.~Starykh  are gratefully
acknowledged. The work was supported by the Department of Energy,
Office of Basic Energy Sciences, Grant No.~DE-FG02-06ER46313.


\begin{references}
\bibitem{valleys} Here we take into account $K$ and $K'$ valley
degeneracy with the extra factor $2$. Considering this degeneracy
explicitly in the Hamiltonian Eq.~(\ref{LEHam}) leads to the same
conclusion.
\end{references}
\end{document}